\documentclass[a4paper]{article}

\usepackage{icrc2013}
\usepackage[english]{babel} 
\usepackage{amsmath}

\title{Geomagnetic field and altitude effects on the performance of future IACT arrays}

\shorttitle{Geomagnetic and altitude effects on IACT arrays}

\authors{
M.~Szanecki$^{1}$,
K.~Bernl\"ohr$^{2}$,
D.~Sobczy\'nska$^{1}$,
A.~Nied\'zwiecki$^{1}$,
J.~Sitarek$^{1,3}$ and
W.~Bednarek$^{1}$
}

\afiliations{
$^1$ Department of Astrophysics, University of {\L}\'od\'z, Pomorska 149/153, PL--90--236 {\L}\'od\'z, Poland \\
$^2$ Max-Planck-Institut f\"ur Kernphysik, P.O.~Box 103980, D 69029 Heidelberg, Germany\\
\scriptsize{
$^{3}$ now at: IFAE, Edifici Cn, Campus UAB, E-08193 Bellaterra, Spain.
}
}

\email{sobczynska.dorota@gmail.com}

\abstract{The performance of IACT's arrays is sensitive to the altitude and geomagnetic field (GF) of the observatory site. Both effects play important role in the region of the sub-TeV gamma-ray measurements.  
We investigate the influence of GF on detection rates and the energy thresholds for five possible locations of the future CTA observatory using the Monte Carlo simulations. We conclude that the detection rates of gamma-rays and the energy thresholds of the arrays can be fitted with linear functions of the altitude and the component of the GF perpendicular to the shower axis core. These results can be directly extrapolated for any possible localization of the CTA. In this paper we also show the influence of both geophysical effects on the images of shower and gamma/hadron separation.}


\begin{document}
\maketitle

\section{Introduction}
\label{sec:intro}
Imaging Air Cherenkov Telescopes (IACT) detect gamma rays using the Cherenkov images of their electromagnetic showers developing in the atmosphere. The IACT technique has rapidly advanced over the last 20 years (see, e.g., a review in  \cite{bib:buckley08}) and, with the current generation of IACT instruments \cite{bib:Aleksic11,bib:Hofmann00,bib:Veritas08}, it is now the most accurate and sensitive detection technique in the very high energy gamma-ray astronomy.
The  Cherenkov Telescope Array (CTA), the next generation of IACT detectors, is expected to improve the sensitivity of present observatories by an order of magnitude, covering the energy range from a few tens of GeV to hundreds of TeV \cite{bib:cta}.

In this paper we study the geomagnetic field (GF) effect, disturbing the IACT technique, which may set an inherent limit on the performance of CTA, especially at low energies. Charged particles in atmospheric showers are deflected by the Earth's magnetic field, which changes the geometry of the light pool\footnote{Defined as the area on the ground with nearly constant density of Cherenkov photons.} (as directions of photons from e$^{+}$ and e$^{-}$ are deflected in opposite directions) and also leads to distortions of shower images. We thoroughly investigate the first, geometrical effect, i.e.\ the influence of GF on the probability of registering a gamma ray due to changes in Cherenkov photon density on the ground. We also quantify the changes of image parameters crucial for the gamma/hadron separation and the direction reconstruction, which allows us to discuss trends in the quality of the separation and reconstruction procedures for the changing magnitude of the GF effect.

We focus here on the subarrays of several large telescopes of CTA, which are dedicated for high sensitivity observations below 100 GeV. 
Then, our results illustrate directly the impact of the GF effect on the energy threshold of the whole observatory, but qualitatively similar effects may be expected also in other classes of telescopes around their threshold energies.

Apart from the local GF at an observatory location, the IACT performance at low energies may be significantly affected by its altitude  (cf.\ \cite{bib:Aharonian01}). 
In this paper we consider five potentially interesting sites for CTA (cf.\ section 11 in \cite{bib:cta}). 
We provide the site-specific information, however, we aim also to derive some more general properties. For each site we study different directions of observation corresponding to different strength of the magnetic field. Then, to disentangle effects due to the altitude and to the GF, for each site we analyse also the IACT performance with vanishing GF. See \cite{bib:Szanecki13} for more details.

\section{GF at the candidate sites}
\label{sec:sites}

Table \ref{tab:sites} gives the geophysical data for the sites considered in this paper. We use the standard parametrization of the GF (see \cite{bib:Campbell03}), $\vec{B}\equiv \left(H,0,Z \right)$, with the $x$-axis pointing to the local magnetic north, the $y$-axis pointing eastward and the $z$-axis oriented downwards.

\begin{figure}[!h]
\centering
\includegraphics[trim = 4.5mm 0mm 0mm 2mm, clip,scale=0.40]{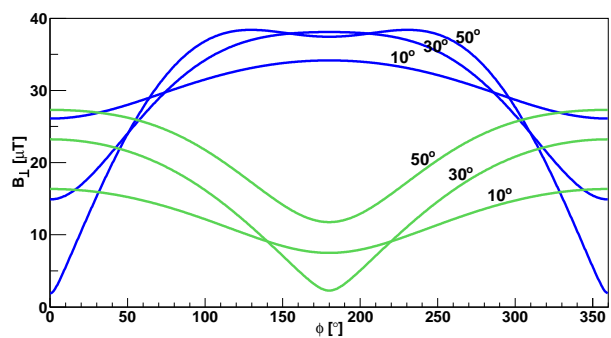}
\caption{Dependence of the transverse component of GF, $B_{\perp}$, on $\phi$ at fixed $\theta$=10$^\circ$, 30$^\circ$, 50$^\circ$ for the H.E.S.S.~(green lines) and Tenerife (blue lines) sites; each line is labeled by the corresponding value of the Zenith angle, $\theta$.}
\label{fig:B_perp}
\end{figure}

Charged particles in the shower observed at the zenith angle $\theta$ and the azimuthal angle $\phi$ are deflected by the  Lorentz force which is proportional to the component of $\vec{B}$ perpendicular to the observation direction $B_{\perp}$ which is a function of the azimuthal angle $\phi$, zenithal angle $\theta$ and local GF components $H$ and $Z$, see Table \ref{tab:sites}. Fig.\ \ref{fig:B_perp} illustrates the dependence of the transverse GF on the observation direction. The change of the sign of the $Z$ component between the southern and northern hemisphere results in an opposite dependence on the azimuthal angle in southern and northern sites.

\begin{table}[!h]
\centering
 \begin{tabular}{|c|c|c|c|c|c|c|c|}
  \hline
  \multicolumn{3}{|c|}{Site} & {{\bf A-S}} &{\bf A-L} & {\bf M} & {\bf S} & { \bf N} \\
  \hline
  \multicolumn{3}{|l|}{Height [km a.s.l.]} & 3.6 & 2.66 & 2.4 & 2.2 & 1.8\\
  \hline
  \multicolumn{8}{|c|}{Local magnetic field $\vec{B}=(H,0,Z)$}\\
  \hline
  \multicolumn{3}{|c|}{$H$ [$\mu$T]}& 21.1 & 20.1 & 25.3 & 30.6 & 12.1\\
  \multicolumn{3}{|c|}{$Z$ [$\mu$T] (+ Down)}& -8.8 & -12.2 & 38.4 & 23.2 & -25.5\\
  \hline
  \multicolumn{8}{|c|}{$B_{\perp}(\theta,\phi)$ [$\mu$T]}\\
  \hline
  \multicolumn{1}{|c|}{$\theta$=$30^{\circ}$}&\multicolumn{2}{|l|}{$\phi$=$0^{\circ}$}& 22.7 & 23.5 & 2.7 & 14.9 & 23.8\\
  \multicolumn{1}{|c|}{$\theta$=$30^{\circ}$}&\multicolumn{2}{|l|}{$\phi$=$180^{\circ}$}& 13.9 & 11.3 & 41.1 & 38.1 & 2.3\\
  \hline
 \end{tabular}
\caption{The geophysical data for Argentina-Salta ({\bf A-S}), Argentina-Leoncito ({\bf A-L}), M\`exico-San Pedro Martir ({\bf M}), Spain-Tenerife ({\bf S}) and Namibia-H.E.S.S. ({\bf N}) sites. The values of $H$, $Z$ and $B_{\perp}$ are obtained from the data given at {\it www.ngdc.noaa.gov/geomag}.}
 \label{tab:sites}
\end{table}

For our simulations we choose $\theta = 30^{\circ}$, which allows to study large ranges of the strength of the transverse GF by changing $\phi$ and moreover is critical for observations in the low energy range.

\section{MC simulations}
\label{sec:mc}
We use CTA Monte Carlo tools to simulate the development of both gamma-ray and proton showers and telescope response. Afterwards, we apply image-analysis procedures to the simulated data.

The EAS (Extensive Air Shower) simulations are performed using the {\fontfamily{pcr}\selectfont CORSIKA 6.98} code adapted for CTA \cite{bib:Heck98,bib:Bernlohr08}; the code includes the description of the influence of GF on EAS. We simulate showers induced by primary gamma rays from a point-like source and by protons arriving from a cone with an opening full angle of $14^{\circ}$. The basic parameters used in simulations are given in Table \ref{tab:mc_cor}.
\begin{table}[t]
\centering
  \begin{tabular}{|c||c|c|}
  \hline
  {\fontfamily{pcr}\selectfont CORSIKA} input & \multicolumn{2}{|c|}{Input value}\\
  \cline{1-3}
  \hline \hline
  Primary particle & Gamma-ray & Proton\\
  \hline
  Energy range & 3--1000 GeV & 10--3000 GeV\\
  \hline
  Power-law index $\Gamma$ & 2.0/2.6 & 2.0/2.73\\
  (simulated/weighted) & &\\
  \hline
  Impact parameter &0--800 m & 0--1500 m\\
  \hline
  Zenith angle & $30^{\circ}$ & 23--$37^{\circ}$\\
  \hline
  \end{tabular}
\caption{Basic parameters used in {\fontfamily{pcr}\selectfont CORSIKA} simulations.}
\label{tab:mc_cor}
\end{table}
At each site we consider the azimuthal angles $\phi=0^{\circ}$ and $180^{\circ}$, corresponding to the minimum and maximum value of $B_{\perp}$ (for $\theta = 30^{\circ}$). We also make a more detailed study for one specific site, in Namibia, by considering seven azimuthal angles between $\phi=0^{\circ}$ and $180^{\circ}$ with uniform steps of $30^{\circ}$. For each site we consider also the case of vanishing GF, to illustrate effects due to the change of the altitude.

\begin{figure}[t]
\centering
\includegraphics[trim = 20mm 40mm 0mm 1mm, clip,scale=0.30]{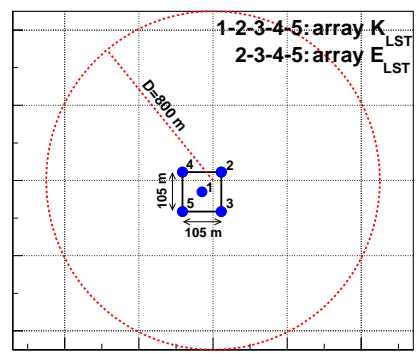}
\caption{Geometric layout of telescopes used in our simulations of gamma rays. The circle shows the simulated area defined by the maximum impact parameter, $D$.}
\label{fig:arrays}
\end{figure}

For the simulations of the telescope response we use the CTA {\fontfamily{pcr}\selectfont sim\_telarray} software \cite{bib:Bernlohr08, bib:BernlohrCTA08} with the default parameter set of the telescope and camera systems assumed in the {\em production-1}: first CTA MC mass production (see, Fig.~18 and Chapt.~8 of \cite{bib:cta} or Chapt.\ 6 of \cite{bib:Bernlohr12}). The {\em production-1} parameters crucial for our simulations are given in Table \ref{tab:mc_tel}.
\begin{table}[t]
\centering
 \begin{tabular}{|c||c||}
  \hline
  {\fontfamily{pcr}\selectfont Sim\_telarray} input & Input value\\
  \cline{1-2}
  \hline \hline
  Telescope type & LST (Type 1)\\
  \hline
  Dish diameter & ${\rm d}=24 \;{\rm m}$\\
  \hline
  Focal length/diameter & ${\rm f/d}=1.3$\\
 \hline
  Camera  Field of view &  ${\rm FoV}=5^{\circ}$\\
  \hline
  Pixel size & $0.09^{\circ}$\\
  \hline
  Photomultipliers & bi-alkali\\
  quantum efficiency &  ${\rm QE}_{\rm peak}=25.7\%$\\
  \hline
  Telescope trigger & Min.\ 4 pe in each of\\
  threshold level & 3 neighboring pixels\\
 \hline
  Min.\ trigger multiplicity & 2 telescopes \\
  \hline
 \end{tabular}
 \caption{Parameters assumed in {\fontfamily{pcr}\selectfont sim\_telarray}.}
 \label{tab:mc_tel}
\end{table}
As an example, we have investigated various trigger configurations, with different number of triggered telescopes, $N$, required to register an event. However, we found that there are no noticeable differences between our results for different $N$, especially concerning the GF effect, apart from the obvious property of systematically higher energy thresholds and smaller detection rates for systems with higher $N$. Then, in this paper we present only our results for $N=2$. We consider two telescope layouts presented in Fig.~\ref{fig:arrays}, namely arrays ${\rm E}_{\rm LST}$ and ${\rm K}_{\rm LST}$ with 4 and 5 Large Size Telescopes (LST), respectively.

We analyse the images obtained from {\fontfamily{pcr}\selectfont sim\_telarray} simulations using the {\fontfamily{pcr}\selectfont read\_cta} program (which is an internal component of the CTA simulation package). After the image cleaning, for which we use the tail-cuts of 5.5/11 photo-electrons (pe), we get the image parameters discussed in Sec.\  \ref{sec:images}.

\section{{Results}}
\label{sec:results}
\subsection{Energy thresholds and trigger rates}
\label{sec:4.1}
In this section we present basic {\it trigger level} parameters, which can be used to describe the performance of ground-based gamma ray detectors, cf.\ \cite{bib:hinton09}.

\begin{figure}[t]
\centering
 \scalebox{0.92}{%
  \includegraphics[width=4.5cm,height=4.6cm]{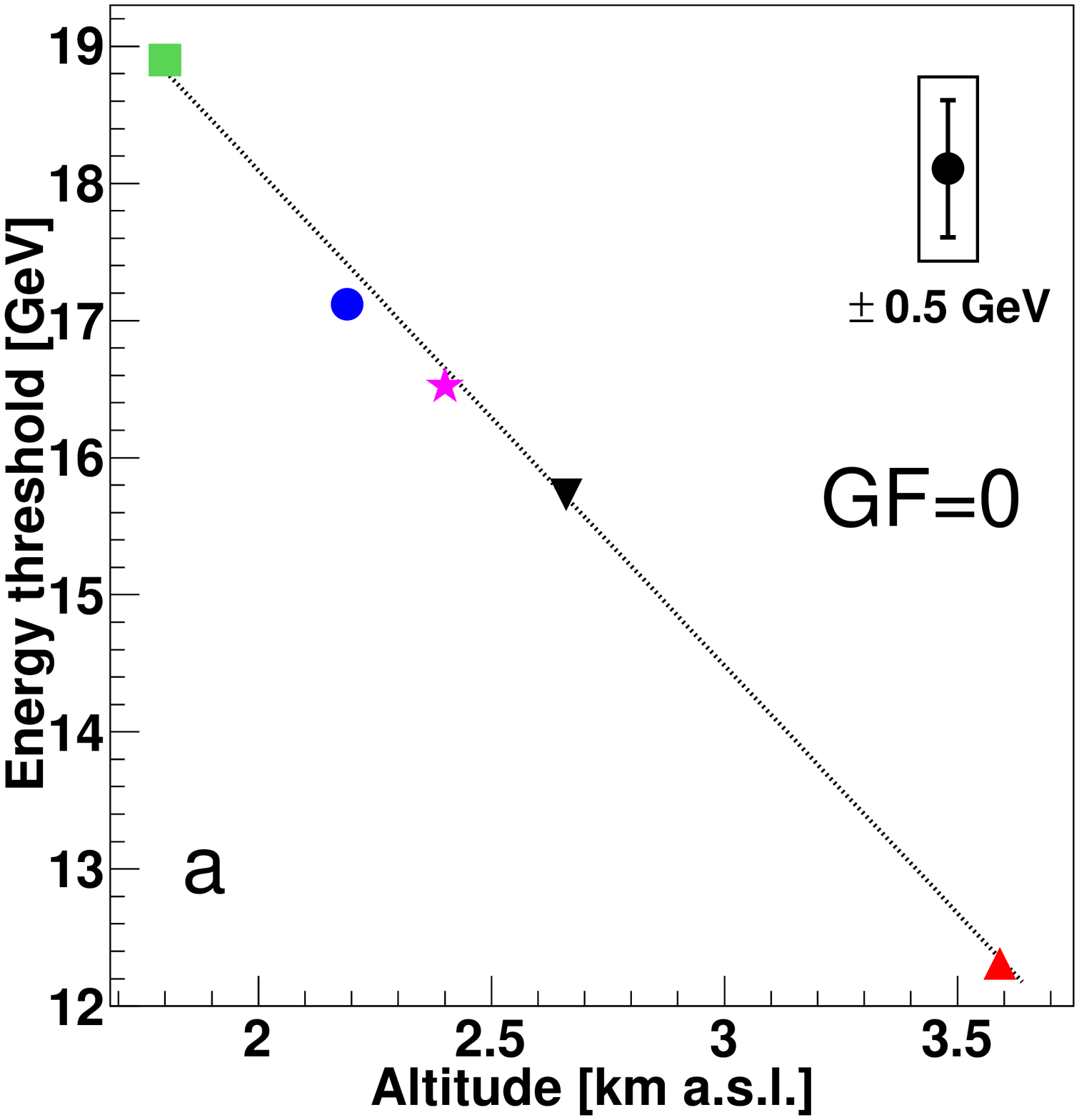}}%
 \scalebox{0.92}{%
  \includegraphics[width=4.5cm,height=4.6cm]{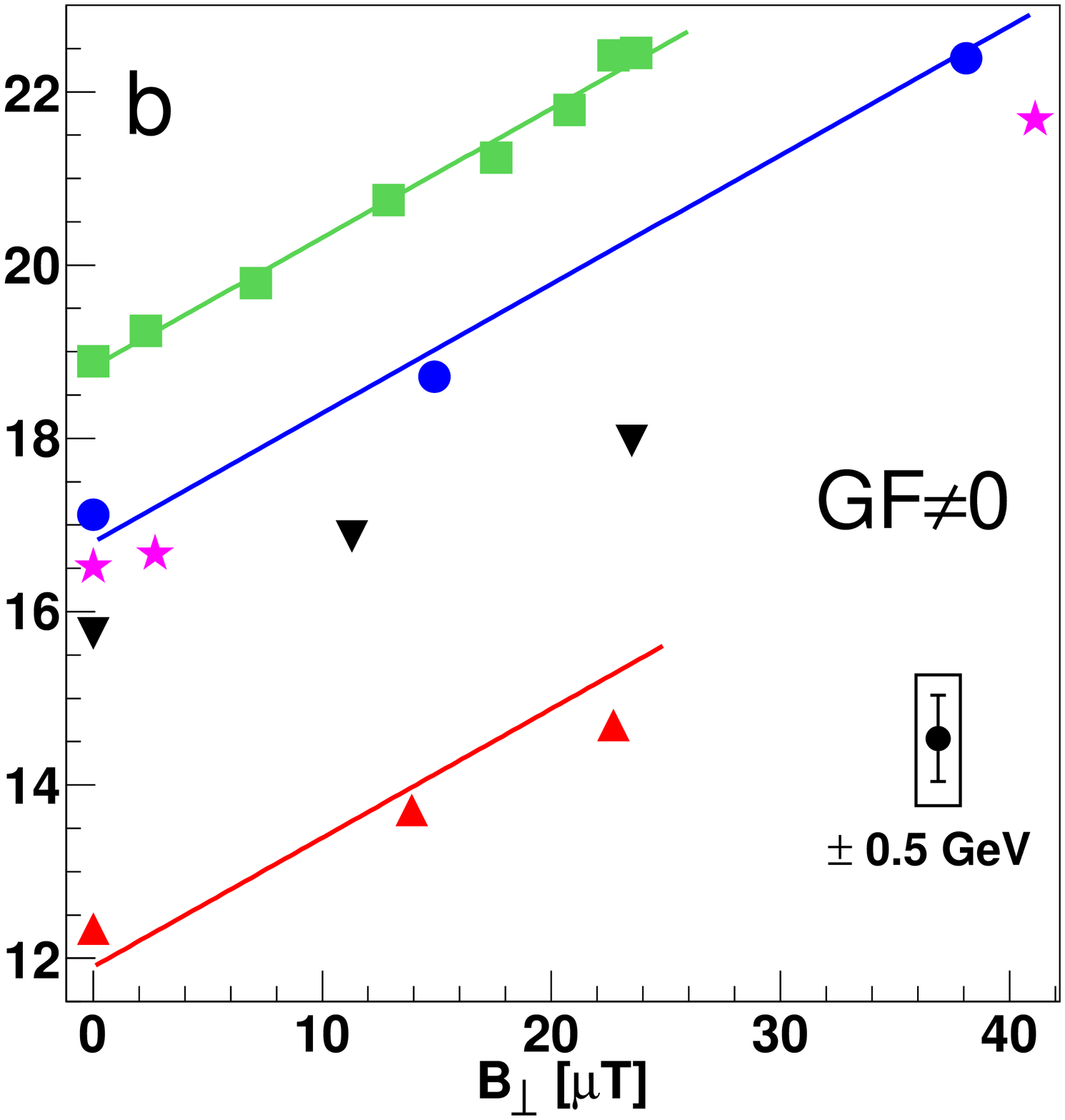}}%
\caption{Energy threshold at the sites with neglected GF (a) (the thresholds differ only due to the difference of altitudes between the sites) and with GF as a function of $B_{\perp}$ (b). The points correspond to Salta (red upward triangles), Leoncito (black downward triangles), San Pedro Martir (magenta stars), Tenerife (blue circles) and Namibia (green squares).}
\label{fig:et_h}
\end{figure}

To illustrate the altitude effect (cf.\ \cite{bib:Aharonian01,bib:Konopelko04}), we show in Fig.\ \ref{fig:et_h}a the energy thresholds for the sites with neglected GF. The energy threshold decrease with increasing altitude due to two effects. First, the atmospheric transmission for Cherenkov light is higher, as the air mass between the shower maximum and the telescope is lower, at higher altitudes. Second, the light pool area is smaller (an obvious geometrical effect) and, hence, the Cherenkov photons density at distances close to the core axis is higher at higher altitudes. These two effects are important only for low energy showers, which produce  Cherenkov light intensities not exceeding significantly the threshold level for triggering a telescope. For the telescope parameters and sites assumed in this work, the dependence of the energy threshold on the altitude of the array can be well described by a linear function, see Fig.\ \ref{fig:et_h}a. Extrapolating it to higher altitudes we get $E_{\rm th} \approx 7$ GeV at 5 km a.s.l., in approximate agreement with \cite{bib:Aharonian01}.

\begin{figure}[t]
\centering
\includegraphics[trim = 8mm 2mm 0mm 1mm, clip,scale=0.40]{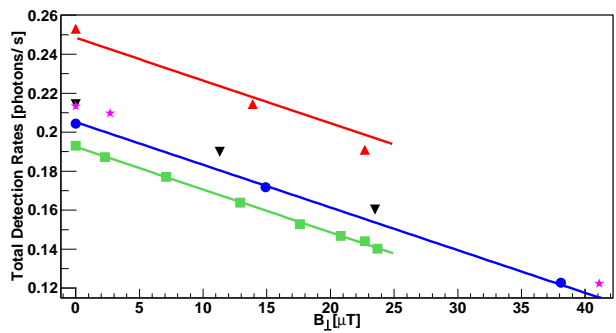}
\caption{The total gamma trigger rates for a pure power-law, ${E}^{-2.6}$ spectrum with flux at 1 TeV equal to 1\% of the Crab source as a function of $B_{\perp}$. The correspondence between markers and sites is the same as in Fig.~\ref{fig:et_h}.}
\label{fig:et_tr_b}
\end{figure}

Fig.\ \ref{fig:et_h}b summarizes our results on the energy threshold for all cases calculated in our work (including the cases of null $B$). Fig.\ \ref{fig:et_tr_b} shows a similar summary for the total trigger rate, $R_{\rm tot}$, obtained by integrating $R(E)$ over the total energy range range (from 3 GeV to 1 TeV). Both quantities are shown as functions of $B_{\perp}$. Remarkably, they appear to scale linearly with ${B}_{\perp}$ at a fixed altitude.

We note that the values of $R_{\rm tot}$ and $E_{\rm th}$ for array ${\rm K}_{\rm LST}$ (with 5 telescopes) are systematically different from those presented in Figs  \ref{fig:et_h} and \ref{fig:et_tr_b} for array ${\rm E}_{\rm LST}$. Specifically, $R_{\rm tot}$ is larger by a factor of 1.2, however, the slopes of $R_{\rm tot}(B_{\perp})$ and $E_{\rm th}(B_{\perp})$ are the same as for array ${\rm E}_{\rm LST}$.

\subsection{Image parameters and $\gamma$/hadron separation}
\label{sec:images}
The separation of the gamma-ray signal from the dominating hadronic background is a fundamental issue in the IACT technique. An effective way to separate gamma rays from the background exploits the differences in the distributions of the image parameters for gamma and hadronic showers. Those differences are used by most of gamma/hadron separation techniques, like scaled cuts and Random Forest (e.g., \cite{bib:albert08b,bib:konopelko99}).

The Hillas width and length parameters \cite{bib:Hillas85}, defined as the second central moments calculated along the minor and major axes of the image, are good separation parameters, at least for large image sizes\footnote{The size is defined as the integrated light of the shower image after image cleaning.}. Namely, the mean width and length as functions of size (referred to as the width and length profile) are different for gamma rays and hadrons.

\begin{figure}[t]
\centering
 \scalebox{0.92}{%
  \includegraphics[width=4.5cm,height=4.6cm]{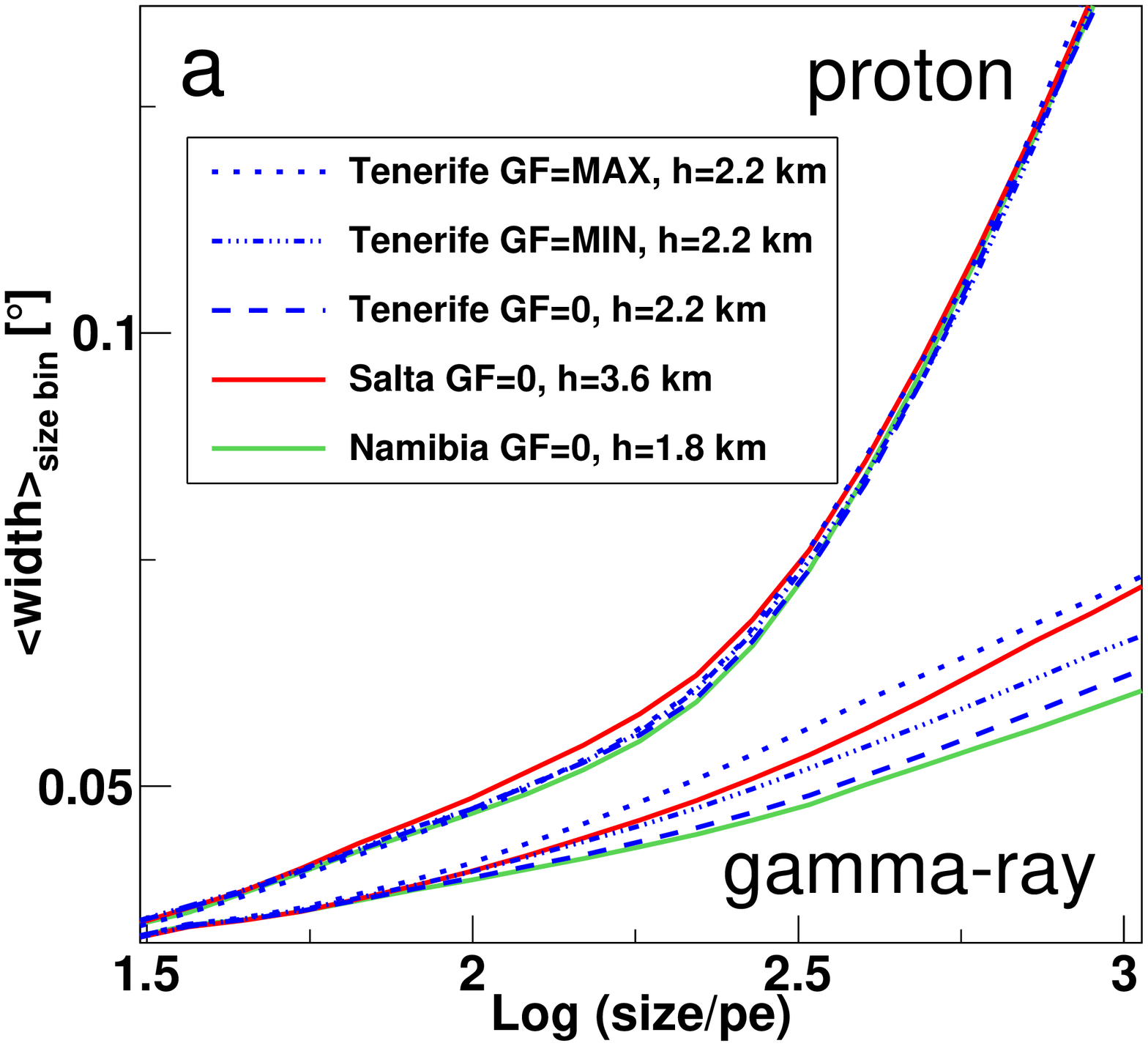}}%
 \scalebox{0.92}{%
  \includegraphics[width=4.5cm,height=4.6cm]{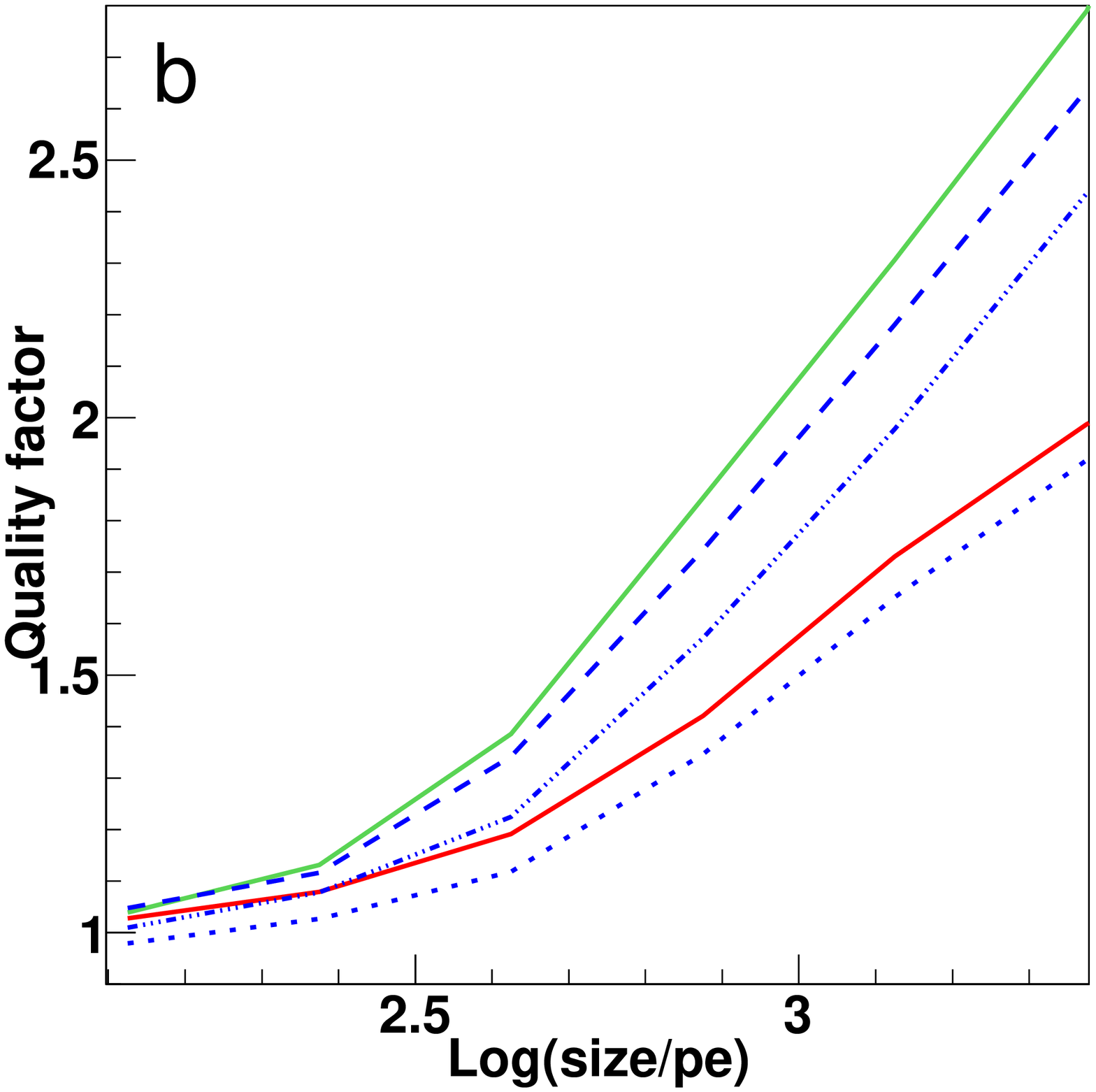}}%
\caption{The mean width for gamma rays and protons (a) and the quality factor for the scaled width cut (b) as the function of size for Namibia (green line), Tenerife (dashed line) and Salta (red line) sites without GF and for Tenerife at $\phi = 0^{\circ}$ (triple-dot-dashed line) and $\phi = 180^{\circ}$ (dotted line) with GF. The plots for Salta and Namibia illustrate the effect of changing the altitude only; the graphs for Tenerife illustrate the effect of changing GF at a fixed height.}
\label{fig:ws}
\end{figure}

The broadening of images by the GF effect should be reflected in the length and width distributions or their profiles. Fig.~\ref{fig:ws}a shows the width profiles obtained from our simulated data by dividing the size range into bins and calculating the mean width, $<$width$>$, of the width distribution for each bin. For gamma rays, the increase of $B_{\perp}$ at a fixed $h$ leads to the increase of $<$width$>$ at sizes exceeding 100 pe,  e.g.\ $<$width$>$ increases by $\sim 20\%$ with the increase of $B_{\perp}$ from 0 to $40\; {\rm \mu T}$. The protonic images, being intrinsically broader than the gamma images, are much less affected by both the GF and the altitude - their influence on the protonic width profiles appears insignificant.

Neglecting the GF effect, we find that for gamma rays the mean width, for a given size bin, increases with increasing altitude, see Fig.~\ref{fig:ws}a, which effect results simply from the decreasing distance between the telescope and the shower maximum, see \cite{bib:Konopelko04}.

The mentioned, scaled cuts technique uses the widths (or, similarly, lengths) scaled to values expected for gamma rays. In a given size range, the scaled width distribution is computed as $w_{\rm s}\equiv (width\;-<\!w\!>)/\sigma_{\rm width}$, where $<\!w\!>$ and $\sigma_{\rm width}$ are the mean width and standard deviation. The separation quality is usually measured by the quality factor, $QF\equiv (n^{\gamma}_{\rm cuts}/n^{\gamma}_{\rm tr})/\sqrt{n^{h}_{\rm cuts}/n^{h}_{\rm tr}}$, where $n^{\gamma,h}_{\rm tr}$ and $n^{\gamma,h}_{\rm cuts}$ are the number of gamma (hadron) events that passed the image cleaning and a given separation procedure, respectively. Given the difference in the gamma and hadron {\it width} distributions illustrated in Fig.~\ref{fig:ws}a, even a simple cut, e.g.\ accepting only events with $w_{\rm s}<1$, results in efficient rejection of hadrons (except for small sizes). Fig.~\ref{fig:ws}b shows values of the quality factor for the $w_{\rm s}$ cut, with $w_{\rm s}<1$, at Salta, Namibia and Tenerife sites. At small sizes, the separation quality is poor  and  differences between different altitudes or GF strengths are small. However, the differences increase with increasing size (and quality factor); we note that both the increase of the altitude from 1.8 to 3.6 km a.s.l. and the increase of $B_{\perp}$ from 0 to $40 \;{\rm \mu T}$ results in a similar reduction of the $QF$--as could be expected from magnitudes of effects illustrated in Fig.~\ref{fig:ws}a.

We check that by introducing the direction cut in ${\mit \Theta}$ image parameter, defined as the angular distance between the simulated and the reconstructed directions of the primary gamma ray, we get a much better separation. For example, accepting only events with ${\mit \Theta}^2 < 0.05$ deg$^2$,  we get $QF > 5$ even for small sizes.

\subsection{Angular resolution}
\label{sec:angular_res}
As pointed out in previous studies, the reconstruction accuracy is affected by both the GF strength \cite{bib:Commichau08}  and the altitude \cite{bib:Konopelko04}. The former effect involves the change of the orientation of the individual shower images by the GF, which obviously spoils the direction reconstruction.

\begin{figure}[t]
\centering
\includegraphics[trim = 10mm 3mm 3mm 2mm, clip,scale=0.40]{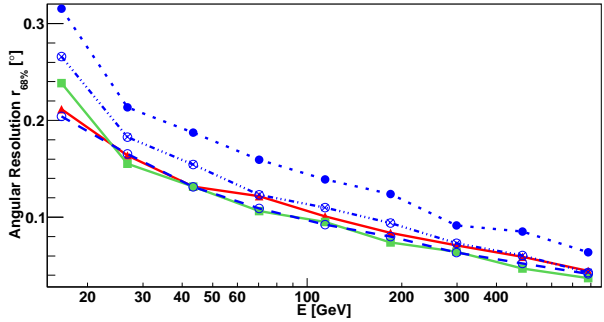}
\caption{Angular resolution as a function of energy for the 68\% containment radius. Three sites are presented: Argentina-Salta without GF (red, solid line), Namibia without GF (green, solid line), Tenerife without GF (blue, dashed line), Tenerife with GF at $\phi = 0^{\circ}$ (blue, triple-dot-dashed line) and Tenerife with GF at  $\phi = 180^{\circ}$ (blue, dotted line). The energy-dependent cuts in {\it Size} are used (e.g.~85, 300 and 2000 pe (per telescope) for 20, 300 and 500 GeV, respectively).}
\label{fig:ang_res}
\end{figure}

The angular resolution ${\mit \Theta}_{68\%}$, containing the fraction of $68\%$, of events in the distribution of the reconstructed event directions, is presented in Fig.~\ref{fig:ang_res}. The increase of $B_{\perp}$ by 40 ${\rm \mu T}$ results in a very strong degradation of the angular resolution, by a factor of three stronger than  the degradation resulting from the increase of the altitude by 1.8 km. 
The influence of the geophysical parameters on the angular resolution is slightly weaker at higher energies, however, the change is small. E.g.\ above the primary energy of 200 GeV the effects are only by $\approx 10\%$ weaker than in a lowest energy regime.

\section{Summary and discussion}
\label{sec:summary}

We have studied the influence of two geophysical factors, the local GF and altitude, on the low-energy performance of the planned CTA observatories. We quantify the changes of the performance parameters, which may be used to balance the geophysical conditions against other criteria for site selection.
We derive an approximately linear scaling of the trigger-level threshold energy and detection rates with both $B_{\perp}$ and $h$, which is a novel result.

Regarding the altitude effect, the possibility of the detection of $\sim 10$ GeV photons at altitudes of $\sim 5$ km was considered already in \cite{bib:Aharonian01}.

We made a basic estimation of the distortion of image parameters, which are crucial for the gamma/hadron separation and the direction reconstruction. As previously suggested by \cite{bib:Konopelko04}, and confirmed in our calculations, the gains in trigger efficiency with increasing altitude are significantly reduced at the analysis level, as the extra photons have images which are less suitable for the separation. We confirm also the conclusion of \cite{bib:Commichau08} that the rotation of images by the GF may be the major obstacle in the IACT performance. We find that the maximum changes of the GF and altitude considered here (i.e. 40 $\mu T$ and 1.8 km) have similar (in magnitude) effects in the distortion of width profiles and in the efficiency of the scaled width cuts. On the other hand, the GF effect is much stronger (by a factor of $\sim 3$) in the angular resolution and in the related efficiency of the direction cuts. We notice that while the trigger-level effects are important only at low energies (around the threshold), the effects in the hadron rejection efficiency or in the reconstruction quality are significant at all energies.

Among  the five considered sites, only the Argentinian sites have notably  better, trigger-level performance parameters than the remaining sites. In the post-analysis performance the GF effect will be strengthened and the altitude effect weakened, however, we note again that we regard the trigger-level information as more basic because it does not depend on the assumed analysis procedure.

\vspace*{0.5cm}
\footnotesize{{\bf Acknowledgment: }{This work was supported by the NCBiR grant ERA-NET-ASPERA/01/10 and NCN grant UMO-2011/01/M/ST9/01891.}


\begin{thebibliography}{10}

\bibitem{bib:buckley08} J.~{Buckley} et~al., $<$arXiv:0810.0444$>$.

\bibitem{bib:Aleksic11} J.~Aleksi{\'c} et~al., Astropart.~Phys.~35 (2012) 435-448.

\bibitem{bib:Hofmann00} W.~Hofmann et~al., AIP Conf.~Proc.~515 (2000) 500-509.

\bibitem{bib:Veritas08} J.~Holder et~al., AIP Conf.~Proc.~1085 (2008) 657-660.

\bibitem{bib:cta} M.~Actis et~al., Exp.~Astron.~32 (2011) 193-316.

\bibitem{bib:Aharonian01} F.~A.~Aharonian et~al., Astropart.~Phys.~15 (2001) 335-356.

\bibitem{bib:Szanecki13} M.~Szanecki et~al., Astropart.~Phys.~45 (2013) 1-12. 

\bibitem{bib:Campbell03} W.~H.~Campbell, Introduction to Geomagnetic Fields, Cambridge Univ.~Press, 2003.

\bibitem{bib:Heck98} D.~Heck et~al., Tech.~Rep.~FZKA 6019.

\bibitem{bib:Bernlohr08} K.~Bernl\"ohr, Astropart.~Phys.~30 (2008) 149-158.

\bibitem{bib:BernlohrCTA08} K.~Bernl\"ohr, in: AIP Conf.~Proc.~1085 (2008) 874-877.

\bibitem{bib:Bernlohr12} K.~Bernl\"ohr et~al., Astropart.~Phys.~43 (2013) 171-188.

\bibitem{bib:hinton09} J.~A.~Hinton and W.~Hofmann, Ann.~Rev.~Astronom.~Astrophys.~47 (2009) 523-565.

\bibitem{bib:Konopelko04} A.~Konopelko, J.~Phys.~G.~30 (2004) 1835-1846.

\bibitem{bib:albert08b} J.~Albert et~al., Nucl.~Instrum.~Methods Phys.~Res.~A 588 (2008) 424-432.

\bibitem{bib:konopelko99} A.~Konopelko et~al., Astropart.~Phys.~10 (1999) 275-289.

\bibitem{bib:Hillas85} A.~M. Hillas, in: Int.~Cosmic-Ray Conf.~(La Jolla) Proc.~3 (1985) 445-448.

\bibitem{bib:Commichau08} S.~C. Commichau et~al., Nucl.~Instrum.~Methods Phys.~Res.~A 595 (2008) 572-586.

\end{thebibliography}
\end{document}